\documentclass{article}
\usepackage{amsmath}
\usepackage{epsf}
\usepackage[dvips]{graphicx}
\usepackage{pazhEref}

\def\etal{{et~al.}}

\begin{document}
\baselineskip 15pt

  \title{\bf MODELING OF THE INTERACTION OF GRB PROMPT EMISSION WITH THE CIRCUMBURST MEDIUM.}
  \author{\bf \hspace{-1.3cm}\copyright\, 2010. \\   D.A. Badjin\affilmark{1*}, S.I. Blinnikov\affilmark{2}, K.A. Postnov\affilmark{1}}

  \affil{
  {\it Sternberg Astronomical Institute, Moscow, Russia}$^1$\\
  {\it Institute for Theoretical and Experimental Physics, Moscow, Russia} $^2$ \\
    }
  \vspace{2mm}

  \sloppypar
  \vspace{2mm}
  \noindent
   We present methodology  and results of numerical modeling of the interaction of GRB prompt emission with the circumburst medium using a modified version of the multi-group radiation hydrocode STELLA. The modification includes the nonstationary photoionization, the photoionization heating and the Compton heating along with the hydrodynamics and radiation transfer.  The lightcurves and spectra of the outcoming gamma-ray, X-ray and optical emission are calculated for a set of models (shells) of the circumburst environment, which differ in dimensions, density, density profile, composition, temperature. In some cases total bolometric and optical luminosities can reach $10^{47}$ and $10^{43}$ erg s$^{-1}$ respectively. These effects can be responsible for irregularities which are seen on lightcurves of some GRB's
X-ray and optical afterglows.

\noindent


\noindent\rule{5cm}{0.5pt}\\
{$^*$ e-mail $<$badjinda@gmail.com$>$}\\
\noindent\rule{16cm}{0.5pt}\\
\vspace{1cm}
 \section*{INTRODUCTION}
 \noindent

Optical (van Paradijs et al. 1997) and X-ray (Costa et al. 1997) afterglows of gamma-ray bursts (GRBs) had influenced and still significantly affect the evolution of our understanding of the nature of GRBs (see e.g. reviews Postnov 1999; M\'{e}sz\'{a}ros 2002).

It is now commonly accepted that the powerful energy release of the GRB itself is associated with gravitational energy transformation during the massive star's core collapse into a black hole or into a neutron star, and the broadband afterglow is believed to arise from interaction of relativistic ejecta (from the GRB central engine) with the surrounding medium (see e.g. Piran 2004). Also observations suggest that some fraction of long GRB-s is connected to unusually bright supernovae (see Woosley \& Bloom 2006 and references there).

Optical afterglow variability at different times  and on the different timescales is sometimes detected. In some cases (GRB 041219A, 050820A, 080319B) it was observed at the same time as the prompt emission, but more often it exposes itself as irregular deviations from the power-law fading of the optical flux at times about $10^4-10^5$ s since the GRB begins (GRB 021004, 030329 etc.). These irregularities could be explained by an interaction between expanding fireball and inhomogeneous outer medium (Lazzati et al. 2002) or by an additional energy injection (Bj\"{o}rnsson et al. 2004).

Obviously, structures of circumburst matter may variate strongly , and the interaction of the prompt emission with these structures may cause transient phenomena lasting up to several years.  (Bisnovatyi-Kogan \& Timokhin 1997, Barkov \& Bisnovatyi-Kogan 2005, Postnov et al. 2004).

Having at our disposal a sample of 58 GRB-s with known redshifts and optical afterglows (Badjin et al. 2009) we had performed a search for transient features in  their optical lightcurves. A notable part of objects was found to have irregularities, i.e. deviations from power-law fading, which looked like temporary rebrightenings or decrease of the fading rate. The table 1 and the figure 1 present typical parameters and the shape of the irregularities found. Unfortunately we were unable to obtain data of X-ray observations for the times when the irregularities were detected for the GRB-s listed in the table 1, but it is well known that X-ray afterglows usually demonstrate even more complicated temporal structure than optical ones (see e.g. Gehrels et al. 2009).

Such effects (in the optical band) can be explained by the radiation of some additional energy either due to the ``Late-Jet -- Cocoon interaction'' (Shen et al. 2009), or due to cooling of the matter previously heated by the main prompt emission. The characteristic times and variability scales in the former scenario seem to be significantly shorter than in the latter one. Obviously both scenarios can take place, but some evidences of the second one can be indicated, such as X-ray spectral lines detection (Postnov et al. 2004 consider GRB 011211, and Gehrels et al. 2009 also discuss some other cases) and presence of the ``plateau'' at late stages of the optical afterglow lightcurves. This article is dedicated to the studying of heating, cooling and radiative processes of the circumstellar matter being illuminated by  the GRB prompt emission.

Barkov and Bisnovatyi-Kogan in their work (Barkov \& Bisnovatyi-Kogan 2005) considered a similar problem of single delta-like pulse Compton heating of an extended optically thin molecular cloud in terms of 2-dimensional hydrodynamics. Unlike their work, in our simulations we did not make any assumptions about the surrounding matter optical depth or about its full ionization state, we also used a continuous prompt emission luminosity lightcurve, took the photoionization heating into account in addition to the Compton scattering, and modeled the resulting thermal radiation by means of the transfer equation solving. Because of significant conceptual and methodological differences we recommend to read their paper separately, as an example of another approach, and in what follows we shall
not discuss the results of Barkov \& Bisnovatyi-Kogan 2005 or compare them with ours.

\section*{DEFINITION OF THE PROBLEM}

As the basic concept of our model we consider a relatively dense shell around the GRB progenitor (the  shell may be previously released by pre-GRB star some time before the explosion), which is being heated up to high temperatures, by the Compton scattering and by the photoionization, and then radiating its thermal energy. The goal is to calculate spectra and lightcurves of this radiation, taking into account non-stationary heating-cooling and ionization processes, radiative transport and hydrodynamics.

Making the assumption that optical luminosity isotropic equivalent is to be about 10$^{43}$ erg s$^{-1}$ and cooling function $\lambda\sim10^{-21}$ erg cm$^3$ s$^{-1}$, one can estimate the radiating volume and obtain the limits  on the shell geometrical parameters: $V \sim \pi\theta_{jet}^2R^2h \sim 3\times10^{43}n_{11}^{-2}\eta_{opt}^{-1}$ cm$^3 $, where $\theta_{jet}$ is gamma-ray collimation angle, $R$ -- the shell average radius, $h$ -- its thickness, $n_{11}$ -- particle density in $10^{11}$ cm$^{-3}$ and $\eta_{opt}$ is the relative effectiveness of the optical emission. The $\theta_{jet}$ angle must be of order of several degrees, and $R \sim10^{15}-10^{16}$ cm, i.e. shell must be distant enough from the GRB source to let the prompt emission arise, but not too far to keep the photon density high enough to heat the medium significantly. The needed shell thickness $h$ is therefore defined mainly by the optical efficiency. In our calculations we use initial $h\sim10^{13}\div10^{16}$ cm, $n\sim 10^{6}-10^{11}$ cm$^{-3}$ and the density profiles both the uniform one and the windlike one ($\rho\propto r^{-2}$).

The question  of how such shells can be formed lies outside of scope  of this work, but there is some evidence  indicating that very massive stars can release a huge amount of mass (up to several M$_{\odot}$) due to pulsational instabilities (Woosley et al. 2007). As the released shells interact with each other or with the outer medium, complicated structures can be formed with the different densities and the density profiles such as thin dense ``walls''.

\section*{THE METHODOLOGY AND THE CODE}

\subsection*{BASIC PRINCIPLES}

To model the processes of interest numerically we used the STELLA multigroup radiation hydrocode (Blinnikov et al. 1998) modified to reproduce the non-stationary processes of heating and the ionization state setting along with the hydrodynamics and the multiwavelength radiative transfer. The original code STELLA is described in Blinnikov et al. (1998), therefore in this section we will focus only on the modifications made. They are following:

-- we have added into a heat balance equation (2.11 in Blinnikov et al. 1998) a time-dependent term responsible for the gamma-ray heating due to the photoionization and the Compton scattering on free and bound electrons;

-- in those radial zones, where gamma-ray heating occurs, the state of the matter is computed by solving a system of non-stationary ionization balance equation, which takes into account mechanisms of collisional ionization, radiative ionization (the photoionization and ionization due to Compton scattering), radiative and dielectronic recombination under the assumption that all ions are in ground state; in those zones, where heating either has not yet begun or has already ended, the ionization state is still computed in  Boltzmann-Saha approximation (like in the original STELLA code), and a smooth connection of these two approximations is performed just after the heating ends;

-- special program tools  have been added to take into account how the gamma-ray beaming and shell curvature change the lightcurves and spectra.

One-temperature one-dimensional fluid model without magnetic field is adopted to  simplify the calculations. It means that one has to treat the lightcurves at their early stages as averaged on the timescale of electron-ion energy equipartition (which in turn depends on medium parameters and model assumptions about electron-ion collision mechanisms, in our conditions this timescale is to be not greater than few hundred seconds for the Coulomb collisions only). Also in the one-temperature model we underestimate the early stage hard X-ray luminosity, because an amount of high energy electrons occurs to be underestimated significantly (following Sazonov et al. 2003, one can obtain that the electrons can be heated by the Compton scattering to temperatures of tens keV while in the equipartition approach their characteristic energy has to be three orders of magnitude lower).

One-dimensionality leads to underestimation of a sideways expansion of the heated matter of the shell into neighbouring cold regions. In our model, however, we simulate some two- and three-dimensional effects such as fluid velocity redistribution acceleration due to mixing (see appendix B in Blinnikov et al. 1998) and temporal delay due to geometrical curvature. The second effect leads to temporal smoothing of the lightcurves and the spectral evolution, and it weakens the influence of the one-temperature approach on reliability  of our simulations.

These simplifications are significant, but they give us a possibility  to pay more attention to the processes of non-stationary kinetics and radiation transfer, keeping the temporal resolution high (i.e. to use more timesteps without considerable increasing of the total calculation time) and using about hundred photon energy groups.

\subsection*{INTERACTION OF GAMMA-RAYS WITH CIRCUMBURST MEDIUM}

We characterize the gamma-ray emission with the  following parameters: spectral boundaries, a peak luminosity isotropical equivalent, a total duration, a spectral shape function and a tabulated lightcurve.

The effectiveness of the radiative heating is determined in the first place by shell opacity. As the shell zones are being heated their state and opacity are being changed,  affecting significantly the gamma-ray flux propagating towards the next outer zones. That is  why it is necessary to keep an opacity table for each zone and for each gamma-ray phase moment in working memory.

In our calculations we determine the optical depth  of the given zone as
  $$\tau_z(n_z,\epsilon,t_p)=h_z\Sigma_{tot}=h_z \left(\sigma_{KN}(\epsilon) \left( n_{fe}(n_z)+n_{be,c}(n_z,\epsilon)\right)+\Sigma_{pi}(n_z,\epsilon)\right),  \eqno (1) $$
where $n_z$ is the zone number, $h_z$ -- zone's geometrical thickness, $\epsilon$ -- gamma-ray photon energy, $t_p$ -- gamma-ray phase time (time since the first gamma-rays crossed zone's outer boundary), $\sigma_{KN}(\epsilon)$ -- Klein-Nishina cross-section, $n_{fe}$ -- number density of free electrons, $n_{be,c}$ -- number density of bound electrons, which can be knocked out by the Compton scattering of gamma-photons ($\epsilon f_s(\epsilon)>\bar{\chi}_c$, see below), $\Sigma_{pi}$ -- inverse mean free path due to the photoionization process
 $$\Sigma_{pi}(n_z,\epsilon)=\sum_{z,i}(n_{zi}\sigma_{zi}(\epsilon)), \eqno(2)$$
 where $n_{zi}$ is the number density of ions with atomic number $Z$ and with $(Z-i)$  electrons, $\sigma_{zi}(\epsilon)$ -- photoionization cross-sections for ground states of these ions (summarized over all atomic shells), computed using analytical fits from Verner \& Yakovlev (1995) and Verner et al. (1996).

Strictly speaking, the cross-section of the Compton scattering on bound electrons differs from the Klein-Nishina cross-section for photon energies less or near 1 keV  (if we consider astrophysically typical elements), but this difference quickly disappears with the growth of frequency. That is  why it is convenient to use one common Klein-Nishina formula $\sigma_{KN}(\epsilon)$ for all the species instead of many different approximations for every ion and all its shells. The Compton scattering on  bound electrons (the process $i+\gamma\rightarrow i^+ + \gamma' + e^-$) plays an important role in our problem because its cross-section for gamma-rays is several orders of magnitude higher than the photoionization cross-section (i.e. the process with complete photon absorption: $i+\gamma\rightarrow i^+ + e^-$) for atomic levels with ionization potentials about few keV.

A photon number spectral density  in a given zone is calculated as
  $$n_{\gamma,\epsilon}(\epsilon, h,t_p)=n_{\gamma,\epsilon}(\epsilon, 0,t_p)R_{z,1}^2(t_p)\frac{e^{-h\Sigma_{tot}(n_z,\epsilon,t_p)}}{(R_{z,1}(t_p)+h)^2},  \eqno(3)$$
here $R_{z,1}$ -- zone's inner radius, $h$ -- coordinate ``thickness'' (is counted along a radial axis from $R_{z,1}$), and $n_{\gamma,\epsilon}(\epsilon, 0, t_p)$ -- spectral photon density at the inner boundary of the given zone at the phase time $t_p$. The parameter $n_{\gamma,\epsilon}(\epsilon, 0, t_p)$ is determined as:
{\small $$n_{\gamma,\epsilon}(\epsilon, 0,t_p)=\frac{L_\epsilon k(n_z,\epsilon,t_p) \eta(t_p)}{4\pi R_{z,1}^2c\epsilon}
   =\frac{L_{iso} F_{\epsilon}(\epsilon) \eta(t_p)}{4\pi R_{z,1}^2c\epsilon F_0} \times \exp\left(-\sum_{n_z'<n_z}(h_z'(t_p)\Sigma_{tot}(n_z',\epsilon,t_p))\right), \eqno (4)$$
}
where $L_{iso}$ is a bolometric isotropic luminosity equivalent, $F_{\epsilon}(\epsilon)$ -- the spectral shape function, $\eta(t_p)$ -- gamma-ray lightcurve, $F_0$ -- normalizing factor. The sum of optical depths $\tau_z'(n_z',\epsilon,t_p)$ in the exponent of (4) is calculated over all zones situated under the given one.

These formulae represent the simplest approach to the gamma-ray transfer in which scattered photons are excluded from our consideration. This allows  us to compute the gamma-ray photon field at each time step for about hundred energy points, without dramatic increase  of the calculation time, that  would inevitably happen if we solved the transfer equation for gamma-rays as well as for thermal photons (and thermal radiation Eddington factors now are recalculated only once per several tens of timesteps). The assumption that the gamma-ray photons suffer not more than a single scattering allows us  to limit the $\tau_z'(n_z',\epsilon,t_p)$ array to $t_p$ in the phase time dimension. Thus we somewhat underestimate the energy of radiative heating and its duration, the more significantly the more shell's optical depth exceeds unity.

In the approximation described above it is convenient to average the photon number spectral density over zone's thickness (in the following expression an explicit indication of dependence on $t_p$ and $\epsilon$ is mostly omitted):
{\small $$ \bar{n}_{\gamma,\epsilon} = \frac{n_{\gamma,\epsilon}(0) R_{z,1}^2}{h_z}
                               \int\limits_0^{h_z}\frac{e^{-h\Sigma_{tot}}}{(R_{z,1}+h)^2}\:dh
    \approx \frac{L_{iso} F_{\epsilon}(\epsilon) \eta(t_p) k(\epsilon)}{4\pi R_{z,1}^2c\epsilon F_0}
             \frac{R_{z,2}\Sigma_{tot}-2+(2-(R_{z,1}-h_z)\Sigma_{tot})e^{-h_z\Sigma_{tot}}}{h_z R_{z,2}\Sigma_{tot}}, \eqno(5)$$ }
 here $R_{z,2}=R_{z,1}+h_z$ -- zone's outer radius.

In these terms we can in general form determine a power of heating as:
$$\dot{Q}=cV_z\int\limits_0^{\infty}\bar{n}_{\gamma,\epsilon}(\epsilon)\epsilon\sum_{\alpha}(n_{\alpha}\sigma_{\alpha}\eta_{\alpha})d\epsilon, \eqno (6)$$
where $V_z$ is zone's volume, the sum over $\alpha$ denotes summarizing over all the processes of interaction of the gamma-rays with the matter, $n_{\alpha}$ -- number density of particles participating in that processes, $\sigma_{\alpha}$ -- process cross-section, $\eta_{\alpha}$ -- a photon energy fraction  that transforms to a knocked out electron kinetic energy.

The STELLA code uses a specific heating power in units $10^{12}$ erg s$^{-1}$ g$^{-1}$ determined as
{\small
\begin{multline*}
 \dot{q}= \frac{10^{-12}c}{\rho_z} \int\limits_{\epsilon_i}^{\epsilon_f}\bar{n}_{\gamma,\epsilon}(\epsilon)\epsilon \times \\
 \times\left(\left(1-\frac{\bar{\chi}(\epsilon)}{\epsilon}\right)
            \Sigma_{pi}(\epsilon)+\sigma_{KN}(\epsilon)\left(n_{fe}\left(f_s(\epsilon)-\frac{T_e}{m_ec^2}\,g_s(\epsilon) \right)
             + n_{be,c}(\epsilon)\left( f_s(\epsilon)-\frac{\bar{\chi}_c(\epsilon)}{\epsilon} \right) \right) \right) d\epsilon,  \qquad (7)
 \end{multline*} }
where $T_e$ -- an electron temperature (in our case it coincides with the ion one), functions $f_s=f(\frac{\epsilon}{m_ec^2})\frac{\sigma_T}{\sigma_{KN}(\epsilon)}$ and $g_s=g(\frac{\epsilon}{m_ec^2})\frac{\sigma_T}{\sigma_{KN}(\epsilon)}$ designate an averaged photon energy fraction that the photon loses or gains in a single act of the Compton scattering, functions  $f$ and $g$ are taken from Sazonov et al. (2003):
{\small $$f(x)=\frac{3(x-3)(x+1)\ln(1+2x)}{8x^3}+\frac{-10x^4+51x^3+93x^2+51x+9}{4x^2(1+2x)^3} \eqno(8)$$
$$g(x)=\frac{3(3x^2-4x-13)\ln(1+2x)}{16x^3}+\frac{-216x^6+476x^5+2066x^4+2429x^3+1353x^2+363x+39}{8x^2(1+2x)^5} \eqno(9)$$ }

In the equation (7) $\bar{\chi}(\epsilon)$ and $\bar{\chi}_c(\epsilon)$ are effective ionization potentials for the photoionization and the ionization due the Compton scattering respectively. The rule of their evaluation is based on the sum over processes (6):

{\small $$ \bar{\chi}(\epsilon) = \frac{\sum\limits_{z,i,n,l:\chi_{zinl}<\epsilon}n_{zi}\sigma_{zinl}(\epsilon)\chi_{zinl}}{\Sigma_{pi}(\epsilon)}, \:
\bar{\chi}_c(\epsilon) = \frac{\sum\limits_{z,i,n,l:\chi_{zinl}<\epsilon f_s}n_{zi}\chi_{zinl}N_{e,zinl}}{\sum\limits_{zi}n_{zi}}. \eqno(10)$$ }

Subscripts $zinl$ designate the $nl$ subshell of the ion with atomic number $z$ and $(z-i)$ electrons, $\chi$ are ionization potentials, $\sigma$ -- photoionization cross-sections (for ionization due to the Compton scattering $\sigma_{KN}$ is used for all $zinl$), $N_{e,zinl}$ -- a number of electrons on the corresponding ion subshell (in the first equation of (10) it is implicitly contained in $\sigma_{zinl}$), and the sum is taken over all ion subshells from which electrons can be knocked out by photons of energy $\epsilon$.

The photoionization rate also depends on the photon number spectral density:
$$P_{zi}=c\int\limits_0^\infty\bar{n}_{\gamma,\epsilon}(\epsilon) \left(\sigma_{KN}(\epsilon)N_{e,zi,c}(\epsilon)+\sigma_{zi}(\epsilon)\right)d\epsilon, \eqno(11)$$
where $N_{e,zi,c}(\epsilon) = \sum\limits_{n,l:\chi_{zinl}<\epsilon f_s}N_{e,zinl}$ is a number of electrons in the ion $zi$ that may be knocked out by photons of energy $\epsilon$ due to the Compton scattering.

\clearpage
\subsection*{THE IONIZATION BALANCE AND THE STATE OF THE MATTER}

 To determine the ionization state of the matter being heated, in the modified code STELLA we solve the following system:
{\small $$\begin{cases}                              \frac{\partial n_{z0}}{\partial t} = -\left(P_{z0}+C_{z0}n_e \right) n_{z0}+\alpha_{z0} n_e n_{z1}, &\text{for neutral atoms;}\\
                       \ldots \\
                       \frac{\partial n_{zi}}{\partial t} =\left(P_{z,i-1}+C_{z,i-1}n_e\right)n_e n_{z,i-1}
                                                           -\left(P_{zi}+(C_{zi}+\alpha_{z,i-1})n_e \right)n_{zi}+\alpha_{zi} n_e n_{z,i+1},
                                                            & \text{for partial ions;}\\
                       \ldots                               &  \text{\hspace{96pt}(12)}\\
                       \frac{\partial n_{zz}}{\partial t} =\left(P_{z,z-1}+C_{z,z-1}n_e\right)n_e n_{z,z-1}
                                                           -\left(P_{zz}+(C_{zz}+\alpha_{z,z-1})n_e \right)n_{zi},
                                                            & \text{for full ions;}\\
                       \ldots \\
                 \sum\limits_{zi} (in_{zi})=n_e, & \text{for electrons;}
         \end{cases}
$$ }
where $C_{zi}$  are collisional ionization rates (initial state is $zi$), $\alpha_{zi}$ -- combined radiative and dielectronic recombination rates (final state is $zi$), $z$ and $i$ possess all values that are possible in a specified composition of the shell. Formulae, tables and subroutines, which are necessary to compute these rates, were taken from the web-site ``Atomic Data for Astrophysics'' http://www.pa.uky.edu/$\sim$verner/atom.html .

We solve the system (12) implicitly at each timestep (until the gamma-rays completely escape the heated zone) by replacement of the derivatives by finite differences $\frac{\partial n_{zi}}{\partial t}=\frac{n_{zi}-n_{zi}^0}{\Delta t}$. The system does  not contain hydrodynamics equations, so all $n_{zi}$ have to be scaled with the total density variation in the zone, which is determined separately. But system's solution affects the heating power (7) and thus provides some kind of feedback on medium hydrodynamics.

As an iteration process convergence criterion  we use zone's gamma-ray opacity (for several energy groups), namely the value $(1-e^{-h_z\Sigma_{tot}})/\Sigma_{tot}$, because during the heating phase it is of our primary interest. If the opacity is changed significantly during the iterative solving of (12), then the photoionization rates (11) are to be recalculated.

Using the system's solution the photon density (of the gamma-ray coming to the neighboring zones), the internal energy density of particles, the pressure and their partial derivatives over the mass density and the temperature are calculated.

Before the heating begins in the zone and some relaxation time (set manually, now it is 10 seconds) after it ends, the state is calculated in the Boltzmann-Saha approximation, like in the original STELLA code. During the relaxation time a smooth connection of the two solutions (the Boltzmann-Saha solution and the system (12), but with $P_{zi}$ excluded) is performed.

\subsection*{INTERACTION OF RELATIVISTIC EJECTA WITH THE CIRCUMBURST MEDIUM}

When studying processes which occur near the GRB central engine it is also important to pay attention to how the circumburst medium is affected by the relativistic ejecta that generates the gamma-rays. If a matter structure is situated at the distance R from the gamma-ray emission source, then the ejecta will reach  it $\Delta t \approx R/(2c\Gamma^2)$ seconds later after the gamma-ray forward front (here $\Gamma$ stands for an ejecta Lorentz-factor). Given $R\approx 10^{16}$ cm and $\Gamma\approx20\div30$ (see Zhang \& MacFadyen 2009), the temporal delay is to be about $\Delta t\approx 180\div400$ seconds.

On the one hand, the non-relativistic STELLA code does  not allow to model such essentially relativistic processes properly, but on the other hand the short delay $\Delta t$ does  not allow to neglect them. We can partially avoid these troubles without complete rewriting  of the code by calculating in the non-relativistic formalism how the matter is penetrated by ``quasi-ejecta'' i.e. an additional shell moving with the speed of light at the distance $c\Delta t$ behind the gamma-ray forward front and having the same isotropic kinetic energy ($M_qc^2/2$) as the realistic relativistic ejecta ($\approx\Gamma mc^2$). In this case the ``quasi-ejecta'' mass $M_q$ plays role of the Lorentz-factor.

Clearly, this replacement is quite crude, that's why we use the ``quasi-ejecta'' approach only to determine a qualitative difference of properties of the matter heated by the radiation and the matter heated both by the radiation and the kinetic shock. In discussion we also describe some features of such modeling of the relativistic ejecta by means of the non-relativistic code.

 \subsection*{OUTPUT PARAMETERS AND RESULT ANALYSIS}

The main studied parameters are luminosities and fluxes in different spectral bands (the bolometric one, U, B, V, R, I, SXR for Soft X-Rays 0.1-2 keV and XR for X-Rays 2-10 keV). A thermal radiation luminosity spectral density is also modeled.

For a terrestrial  observer the flux is equal to
$$F=\frac{L_{iso}}{4\pi D_{phot}^2}, \eqno(13)$$
here $L_{iso}$ is an isotropic luminosity equivalent, and $D_{phot}$ is a photometric distance. The term isotropic equivalent designates  such a luminosity which the object would have if it radiated in all directions the same energy per solid angle as in the direction of the observer's detector ($L_d$), so
$$\frac{L_{iso}}{4\pi}=\frac{L_d}{\Omega_d}, \eqno(14)$$
here $\Omega_d$ is detector's solid angle as it is seen from the source. This is the estimation of $L_{iso}$ that we can obtain from our observations. Thus we need to define its calculational analog to be able to compare our simulation results with the observed parameters correctly.

To compute $L_{iso}$ we have to know a radiation pattern of the source and a brightness distribution over its surface. There are several additional features that complicate the problem: the gamma-ray collimation (which cuts-off a relatively small area of the spherical shell), possible noncoincidence of a jet axis with a line of sight, the temporal delay of the arrival time of the photons coming from different points of the shell caused by shell's curvature ($\delta t\propto (1-\cos(\theta))R/c$) .

If the jet half-opening angle is $\theta_{jet}$, the off-axis angle is $\theta_{oa}$ (see fig.2), and spherical coordinates of the shell points relative to its centre are parametrized as $(\theta,\varphi)$, then
$$L_d(t)=\int\limits_{S}\lambda(t^\prime(\theta,t))\int\limits_{\Omega_d}g(\theta^\prime,\varphi^\prime)d\Omega^\prime dS, \eqno(15)
$$
here $(\theta^\prime,\varphi^\prime)$ angular coordinates relative to a surface element $dS$ centre and to its normal, $\lambda(t^\prime(\theta,t))=L_{tot}(t^\prime(\theta,t))/4\pi R^2(t^\prime(\theta,t))$ is a luminosity surface distribution at a time $t^\prime=t-\delta t(\theta,t)$, $L_{tot}$ -- the full shell luminosity at its outer edge ($R$), a coefficient $g(\theta^\prime,\varphi^\prime)\propto \cos(\theta^\prime)$ sets an angular distribution of the surface element ($dS$) radiation, integral over $dS$ is taken on the shell outer edge, and integral over $d\Omega^\prime$ is taken over detector's solid angle. A normalizing condition for $g$ is $\int\limits_{0}^{2\pi}\int\limits_{0}^{\pi/2}g(\theta^\prime,\varphi^\prime)\sin(\theta^\prime)d\theta^\prime d\varphi^\prime=1,$ and it gives $g(\theta^\prime,\varphi^\prime)=\cos(\theta^\prime)/\pi$. Then, taking into account an infinitesimality of the solid angle $\Omega_d$ (i.e. $g(\theta^\prime,\varphi^\prime)$ are constant inside it), the integral over $d\Omega^\prime$ can be replaced by $\cos(\theta)\Omega_d/\pi$. And taking into account that  $dS=R^2(t^\prime(\theta,t)\sin(\theta)d\theta d\varphi$, one can obtain
$$ L_{iso}(t)=4\pi\frac{L_d(t)}{\Omega_d}=\frac{1}{\pi}
   \int\limits_{0}^{\theta_{jet}+\theta_{oa}}
        L_{tot}(t^\prime(\theta,t))(\varphi_1(\theta)-\varphi_0(\theta))\cos(\theta)\sin(\theta)d\theta, \eqno(16)$$
here $(\varphi_1(\theta)-\varphi_0(\theta))$ indicates what part (in units of angle) of an elementary ring $\sin(\theta)d\theta$  (dashed concentric rings on the left panel of the figure 2) is situated inside a gamma-ray heated spot of the shell, besides $\theta$ this difference depends  also on $\theta_{jet}$ and $\theta_{oa}$.

If the thermal radiation photon emitted from the polar angle $\theta$ is detected simultaneously with the ``on-axis'' photon emitted at time $t$, then emission time $t^\prime(\theta,t)$ (by observer's watch) of the former one can be found from the equation
$$t^\prime+\frac{R(t^\prime)(1-\cos(\theta))}{c}=t. \eqno(17)$$

It is clearly seen that $\cos(\theta)\sin(\theta)d\theta=-d(\cos^2(\theta))/2$ and therefore it is convenient to choose partition over $\theta$ providing a uniform grid of  value $x=\cos^2(\theta)$ points.

The most principal output of photometric calculations with the STELLA code is a luminosity spectral density table $L_{tot,\nu}$ for different time (up to several thousands moments) and frequency points. Calculations using formulae (16)-(17) are carried out for each frequency point to obtain an array of $L_{iso,\nu}(t)$. The luminosity lightcurves in different filters are computed by intergating of $L_{iso,\nu}(t)$ multiplied by different transmission curves over different spectral ranges.

Also the ``red shifted'' array $L_{iso,\nu}(t)$ (time values are also changed in it) is used in calculations of fluxes and stellar magnitudes. A cosmological model with $H_0=73.5$ km s$^{-1}$ Mpc$^{-1}$, $\Omega_\Lambda=0.76$, $\Omega_\rho=0.24$ is used and no supposition about an extinction either in a host galaxy or in the Milky Way is made.

\section*{NUMERICAL MODELING}

Using the methodology described above we carried out a set of calculations for several dozens of shells, which differed in geometrical configuration, density and temperature. Parameters of the most typical of them are presented in a Table 2.

The incoming gamma-ray radiation was set by three FRED-pulses (Fast Rise - Exponential Decay) with a characteristic fading time of 1.3 seconds, a peak isotropic luminosity equivalent 3$\times$10$^{53}$ erg s$^{-1}$ (see fig. 3) and a Band spectral shape function (Band et al. 1993)

$$F_\epsilon(\epsilon)=F_0\times\begin{cases}
\exp(-\epsilon/\epsilon_0)\epsilon^{1-\alpha}, & \epsilon<(\beta-\alpha)\epsilon_0   \text{\hspace{160pt}(18)}\\
((\beta-\alpha)\epsilon_0)^{\beta-\alpha}\exp(\alpha-\beta)\epsilon^{1-\beta}, & \epsilon\geq(\beta-\alpha)\epsilon_0
\end{cases}$$

with $\epsilon_0=300$ keV, $\alpha=0.9$ and $\beta=2.001$ in an energy range 1 keV - 30 MeV. Angles were taken $\theta_{jet}=10^\circ$, $\theta_{oa}=3^\circ$. The peak luminosity was taken a bit higher than that of the real GRB-s with the afterglow irregularities (see the table 1), because it represented the radiation that had not yet been absorbed, therefore it was reasonable to add some luminosity reserve (but not great, so our accepted luminosity was still of typical value for GRBs).

In calculations with the ``quasi-ejecta'' model, we took the mass of the ``quasi-ejecta'' $M_q=0.5M_{\odot}$ which  corresponded to an isotropic kinetic energy $E_{kin,iso}=M_qc^2/2=4.5\times10^{53}$ erg, i.e. to the value that relativistic ejecta should possess at a distance of $\sim10^{16}$ cm from the central progenitor. Initially the ``quasi-ejecta'' shell was placed 200 light-seconds behind the gamma-ray radiation front.

Results of calculations for typical models (see table 2) are presented in figures 4-16.

 \section*{DISCUSSION}
 \noindent
 \subsection*{THE RADIATION-TO-MATTER INTERACTION}

First of all it should be mentioned that the  use of the non-stationary system (12) for the ionization balance calculation leads to a remarkable difference from either fully ionized matter approximation or Boltzmann-Saha solution. The figure 4 shows the dependencies of the optical depth (in gamma-rays) on the photon energy calculated in different approximations. It is seen (fig. 4a), that when the system (12) is used, the matter quickly (much less than 1 second) becomes fully ionized, but at the very early stages, however, it has much higher optical depth (and therefore absorbs much more energy). In the Boltzmann-Saha approximation the matter temperature occurs to be not sufficient to ensure the full ionization and thus the shell optical depth and the absorbed energy are significantly overestimated.

In the figure 5 spectra of the incoming and outcoming gamma-radiation in its different lightcurve phases. A considerable deficiency of low-energy photons at the early stages of the prompt emission is predicted (but due to the cosmological redshift this feature should fall into X-Ray band and probably should not be observed by gamma-ray detectors). It is also seen that outcoming gamma-ray spectra become harder than incoming ones and that luminosities may significantly decrease.

The figure 6 shows temperature radial profiles at different moments during the shell heating for the ``wall'' model with 100 radial zones, i.e. temporal resolution is $\approx17$ s. Sharp peaks seen in profiles for $t=167$s and $t=1670$s correspond to those zones where the heating has just begun.

Geometrical extents and column density of shells are also of great meaning. A combination of these two parameters determines shapes and amplitudes of the lightcurves in a larger measure than the mass density or the total mass of the shell. Depending on the geometrical thickness and the matter column density there can arise shock waves in the shell, coming out from the inner layers that have absorbed more energy. In the ``ball'' model the windlike density profile the inner zones are the most opaque (in spite of that they have smaller mass than inner zones of other models) and they intercept the main part of gamma-radiation, so the shockwave arising there heats the outer zones more effectively than the radiation itself. That is  why the main maximum of a bolometric luminosity lightcurve takes place not just after the heating ends but near the moment of the shock breakout (see fig.7d).

If the inner layer absorption is not so high or the shell has not large thickness (e.g. like ``the wall'' does), then the matter is heated mainly by the gamma-radiation. In such a case the total radiative cooling time appears to be  shorter, the peak luminosity is higher and more energy is radiated in X-Rays for the first few hours (e.g. see fig.7a). In intermediate cases both the cooling after the radiative heating  and the shock breakout are clearly seen.

A temporal behavior of the thermal radiation is also affected by the shell curvature. This leads to a smoothing of the lightcurve on the timescale of $\delta R/c$, where $\delta R$ is a typical variation of distance of an emitting point from the image plane. Actually, ``the wall'' has its total thickness of $5\times10^{13}$cm, but because of the shell curvature the delay of a large angle radiation is about $\delta t=R(1-\cos\theta_{jet})/c\approx5000$  seconds. This effect is clearly seen in the X-Ray lightcurves (figure 8a, figure 9).

 \subsection*{OBSERVATIONAL DETECTION POSSIBILITIES OF THE MODELED EFFECTS}

Along with a development of a computational tool for  numerical modeling, it is important to understand how results of its predictions can be detected observationally. The studied thermal effects are  more likely to expose themselves as lightcurve irregularities against the power law synchrotron afterglow background. To reveal in what models, at which times and in what spectral bands the thermal effects can be observed, we draw in our graphs characteristic regions containing the main part of observed real afterglow points. To plot R-filter afterglow regions (both for luminosities and for fluxes corrected for the Galactic extinction) we took data from a database of the optical afterglow observations used in work Badjin et al. 2009. X-ray afterglow regions in a range 1-30 keV (also corrected for the Galactic extinction) were defined based  on  review of Gehrels et al. 2009.

It follows from a comparison of thermal X-Ray lightcurves (figure 8) and characteristic afterglow regions that the thermal effects in X-rays can show themselves as either rebrightening (fig. 8a) or considerable deceleration of the total flux fading -- both followed by an abrupt growth of the fading rate and then return to the power law (fig. 8b). Characteristic times, when these features appear, are about tens minutes after the GRB prompt emission begins, and characteristic durations are of few hours. Obviously, the dimmer the power law afterglow is, the clearer the thermal effects can be seen. Their luminosity and relative visibility also depends on the shell properties (see the fig.8 and the table 2).  The column density and the matter distribution must provide a relatively uniform profile of the heating power (i.e. not to capture the main part of radiation near the shell inner edge), but on the other hand an enough amount of heated matter must be seen by the observer. Probably models like ``the wall'' are the closest ones to a satisfaction of these opposing requirements.

In the figure 10 there are shown the  calculated lightcurves of near-earth fluxes in ranges 0.1-2 and 2-10 keV. Here no extinction in the Milky Way or in a host galaxy was taken into account. It makes difficult to compare these lightcurves with observed ones properly, so we did not show afterglow characteristic regions for the X-ray fluxes. However it is still possible to obtain ideas about unabsorbed fluxes and their temporal behavior.

The R-band characteristic afterglow regions are drawn in the figures 7 (for luminosities in a proper system), 11 and 12 (for fluxes observed from earth). The optical luminosity lightcurves suggest that the thermal radiation is probably too weak and can be detected only at very late stages when a non-thermal afterglow radiation fades away. But in fluxes the modeled effects seem brighter and expose themselves considerably earlier (cf. the fig.7a and the fig.11a). This is caused by the fact that the synchrotron luminosity spectral density decreases with frequency (near the $10^{15}$ Hz), while the modeled thermal luminosity spectral density grows, thus, after the cosmological redshift, in the same observer's band falls dimmer parts of the synchrotron spectrum and brighter ones of the thermal. This may compensate or even overcome the common decrease of flux with growth of photometric distance. That is  why the probability of the optical detection of the thermal ``irregularities'' increases significantly with redshift. However, the most common way for the thermal radiation to become apparent is the decreasing of the total afterglow fading rate or ``plateau'' at late afterglow stages (after few days since GRB detection), and at the latest stages even ``bumps'' may be produced, if the thermal radiation is bright enough.

It is worth  to be mentioned that because of the redshift the luminosity estimations presented in the table 1 do not correspond to a proper system R-band (but only to that spectral ranges falling into the terrestrial  R-band), and it is not completely correct to compare them with the calculated R-band luminosities directly.

We can also suppose that we should look for observational detections of the thermal effects amongst GRB-s with high redshifts. And it is interesting to note that GRB 050904 (z=6.29) optical afterglow lightcurve do show a certain deviation from the power law resembling our calculated flux lightcurve (for z=6.29) in the ``wall'' model. Similar feature is seen in the lightcurve of the GRB 090423 afterglow (z=8.2).

Besides relatively dense shells the case of a significantly more rarefied  cloud was studied (the ``cloud'' model in the table 2, see also figures 14 and 15). Although its mass is several tens times lower than that of the other shells, its  optical luminosities and fluxes are nearly of the same order. I.e. if the cloud had more radiating volume (with the same density), this would increase its calculational optical luminosity to a value about that in the table 1. The cooling time would be also increased. But such a calculation requires much more radial zones in the grid , and therefore much more computational time, so we have not yet performed it.

\subsection*{THE INFLUENCE OF THE RELATIVISTIC EJECTA. PROSPECTS OF A FURTHER CALCULATION DEVELOPMENT.}

The preceding discussion was concerned with the effects that would take place, if the circumstellar matter structures were heated by the gamma-ray radiation only. This is satisfied either during the first $10^2-10^3$ seconds (for the observer on earth ) since the GRB begins, or in the case when the shell is very large and the relativistic ejecta slow down in its interior taking no significant effect on the main bulk of matter. In other cases the radiatively heated matter will be affected by the ejecta and relativistic or subrelativistic motion will arise in it.

We tried to estimate the influence of the relativistic ejecta on the thermal radiation processes, as far as it was possible to do this with the non-relativistic code. It had occurred that even quite simplified calculations in the ``quasi-ejecta'' approach took much time: the high temporal resolution (less than $10^{-4}$ s) was needed during a long time (more than $10^{4}$ s) because of high relative velocities of matter elements. We avoided this trouble partially by artificially increasing a mixing parameter (see an appendix B in Blinnikov et al. 1998), to increase an effectiveness of a velocity redistribution. But this is quite speculative, because the realistic mixing parameter values are to be determined from 3D-calculations which yet were not carried out. Nevertheless, we believe, we can do some qualitatively correct conclusions.

The figure 16 shows for a purpose of comparison  the luminosity lightcurves in the ``bolometric'' (1-50000 {\AA}), X-ray and optical (the U-filter is taken because it will fall into red and infrared regions due to the redshift) spectral regions for the ``wall'' model and for three different types of heating: the radiation only, the ``quasi-ejecta'' only and the radiation combined with the ``quasi-ejecta'' (but in the last two cases the shell temperature of 3000 K was taken, to let the lightcurves to reflect the heating dynamics more appreciably). As expected, the radiative heating dominates during early stages, and the luminosity due to a kinetic heating becomes equal to the luminosity due to radiative heating nearly after 1500 seconds (for a near outer observer unaffected by redshift) since the prompt emission begins. After that the shock heating luminosity  exceeds the radiative heating luminosity, but not more than three times, and then they are equal to each other again until near $\sim10^4$ seconds the radiative heating energy reserve is  mainly exhausted, and the radiative heating bolometric luminosity tends to fade quickly. Soft X-ray lightcurves demonstrate the similar behavior.

In the 2-10 keV range the emission caused by the radiative heating exceeds the kinetic heating emission considerably (more than 6 times) until it begins  to fade abruptly. On the contrary, in the U-filter kinetic energy radiation dominates nearly all the time (in the fig.16 one should see the difference only between the ``kinetic'' and ``combined'' optical lightcurves, because the ``radiative'' one is calculated for the initial temperature of 12 000 K, and therefore its initial level is nearly $(T_1/T_2)^4=256$ times higher, as it is clearly seen in the fig.16b).

One can conclude that the thermal energy of the gamma-ray conversion is radiated mainly in higher frequencies and for shorter time, than that gained by the interaction with the ``quasi-ejecta''. The relative increase of the radiation duration due to the ``quasi-ejecta'' influence is to be especially pronounced  for geometrically thin shells (the duration increases by an order of magnitude), while for more extensive shells the difference is not so significant (we also carried out the calculations with the ``quasi-ejecta'' and the radiative heating for the ``thin layer'' model). If the shell is large enough, then the ``quasi-ejecta'' will be quickly decelerated in its interior and lag behind the gamma-ray radiation front considerably, and the bolometric lightcurve will exhibit a strong secondary maximum at a few days after burst's beginning. The luminosity spectral density maximum will lay in a soft X-ray or far ultraviolet region at that moment.

Thus, generalizing, we can say that, though the ``quasi-ejecta'' affects the lightcurves significantly, it is still necessary to take into account the effects of the gamma-ray-to-heat conversion. It is also  obvious that the relativistic ejecta colliding with a dense structure (several $M_{\odot}$) should become non-relativistic rapidly. If so, the replacement of its Lorentz-factor by a ``quasi-ejecta'' effective mass will lose its original meaning of a kinetic energy storage, because the Lorentz-factor will to drop to $\Gamma\approx1$, but the ``quasi-ejecta'' mass will hold unchanged. While the low rest mass relativistic ejecta will dissipate its kinetic energy in the first zones and will turn into a relatively weak non-relativistic disturbance (itself, but not the shocked matter), the heavy ``quasi-ejecta'' ($M_q=0.5M_{\odot}$) will still influence the matter significantly. I.e. they strongly differ in a radial profile of an energy exchange effectiveness with the medium. That is why there are some reasons to believe, that the ``quasi-ejecta'' approach changes the lightcurves excessively, with respect to  what could be expected from the real relativistic ejecta, at least in time domain.

Certainly, to describe correctly the processes taking place when the circumstellar matter is affected by the radiation and the relativistic ejecta of the GRB, one needs for calculations in the framework of relativistic hydrodynamics and radiative transfer. This sets quite a number of challenges. We will now point briefly several of them. Firstly, the nature of the ejecta itself is still unclear, e.g. whether it is  dominated either by baryonic load or by an electro-magnetic field. It is crucial for our understanding how the ejecta interacts with the medium. Secondly, a more sophisticated model of medium structures is required. There obviously must be some amount of matter between the central object (the GRB progenitor) and the shell, and this probably should change the outflow dynamics somehow. Also, the shells are likely to have (and, probably, they do) a complicated 3-d structure: e.g. ``windows'' (regions of rarefaction) that may let the gamma-rays to go through relatively unabsorbed, or higher density clumps that intercept an excessive portion of energy of the radiation and the outflow. Thirdly, it is necessary to use more precise approximations for the problems of gamma-ray transfer and particle kinetics (for example it is useful to solve a kinetic equation at least for electrons during the first seconds of heating in the phase time scale). Finally, because the afterglow emission is essentially non-thermal, it is desirable to have possibility to calculate magnetic fields for the purpose of self-consistent modeling of the synchrotron emission and the magnetic field influence on kinetics of particles (e.g. on an energy redistribution time). We believe that the experience described in this work will be useful in a more sophisticated calculational project dedicated to the phenomena taking place in vicinity  of the GRBs.

 \section*{CONCLUSIONS}
  \noindent

The main results of our work are following. Using the radiation-hydrocode STELLA as the basis, the computational tool was developed, that allows, under reasonable assumptions, to model the processes of time-dependent heating and matter state evolution along with the hydrodynamical and radiative transfer processes.

As one of possible applications the problem of radiation of the dense circumstellar medium structures being heated by the GRB prompt emission is considered. The spectra and lightcurves obtained from the calculations allow us  to assert that modeled processes can be responsible, at least in some cases, for the irregularities and plateaus, that are  seen in several GRB optical and X-ray afterglow lightcurves.

The modeling of the relativistic ejecta influence on the features of thermal radiation of the matter, previously heated by gamma-rays, reveals that it is important to pay attention to both these kinds of heating and also sets the problems that must be solved to do this correctly.


\section*{ACKNOWLEDGEMENTS}

The authors are thankful to referees for their useful remarks. The work is partially supported by RFBR grants 10-02-00599-a, 10-02-00249-a, NSh 2977.2008.2, 3884.2008.2,  by the Federal Agency for Science and Innovation contract: 02.740.11.0250 and by the SNSF grant No.~IZ73Z0-128180/1 (the SCOPES programme).


\clearpage
\centerline {\bf TABLES}

\begin{table}[ht]
  \vspace{6mm}
  \centering{{\bf Table 1.} Typical parameters of optical lightcurve deviations from the power law fading. } \label{table 1}

  \vspace{5mm}
    \begin{tabular}{l|c|c|c|c|c|c}
  \hline\hline
   GRB    & z      & $\log L_\gamma^{\mbox{\small a}}$   &$L_{R,peak}^{\mbox{\small b}}$ & $T_{peak}^{\mbox{\small c}}$ & $T_{90}^{\mbox{\small d}}$ & $E_{bump}^{\mbox{\small e}}$  \\
          &        & erg s$^{-1}$                & erg s$^{-1}$        & days  & days   & erg                        \\
  \hline
  000301C & 2.0335 & 52.39$\pm$0.14$_{[15-150]}$ &$1.45\times10^{44}$  & 1.276 & 0.9    & $(6.68\pm1.7)\times10^{48}$ \\
  020124  & 3.198  & 52.72$\pm$0.03              & $3.73\times10^{43}$ & 0.356 & 0.215  & $(1.9\pm0.7)\times10^{50}$ \\
  021004  & 2.3351 & --                          & $5.73\times10^{44}$ & 0.024 & 0.15   & $(1.2\pm0.5)\times10^{49}$ \\
  030328  & 1.52   & 52.33$\pm$0.07              &$1.28\times10^{44}$  & 0.115 & 4.09 ? & $(3.1\pm1.5)\times10^{48}$ \\
  030429X & 2.65   & 53.05$\pm$0.07              &$2.89\times10^{43}$  & 0.751 & 0.68   & $(6.7\pm4)\times10^{48}$ \\
  060206  & 4.048  & 52.49$\pm$0.03              &$4.33\times10^{43}$  & 0.111 & 0.025  & $(6.2\pm2.2)\times10^{48}$ \\
  \hline
    \multicolumn{7}{l}{}\\ [-3mm]
    \multicolumn{7}{p{15cm}}{$^{\mbox{\small a}}$ Maximal luminosity isotropic equivalent in a photon energy range 1 keV -- 10 MeV, but for GRB 000301C (for which only a single power law spectrum was reported)  $L_\gamma$ corresponds to 15 -- 150 keV } \\
    \multicolumn{7}{p{15cm}}{$^{\mbox{\small b}}$ Maximal isotropic optical luminosity deviation from the power law fading in a proper system frequency range corresponding to the earth R-filter.} \\
    \multicolumn{7}{p{15cm}}{$^{\mbox{\small c}}$ Time since GRB beginning (in a proper system timescale) when the maximal deviation occurs.} \\
    \multicolumn{7}{p{15cm}}{$^{\mbox{\small d}}$ Duration of the deviation while 90\% of its energy is being emitted} \\
    \multicolumn{7}{p{15cm}}{$^{\mbox{\small e}}$ Total optical energy of the deviation.} \\
  \end{tabular}
  \end{table}

\begin{table}[h]
  \vspace{6mm}
  \centering{{\bf Table 2.} Initial parameters of the most typical shell models.} \label{table 2}

  \vspace{5mm}
    \begin{tabular}{l|c|c|c|c|c|c|c}
  \hline\hline
   Reference & $R_{in}-R_{out}^{\mbox{\small a}}$  & $M^{\mbox{\small b}}$           & $n_{bar}^{\mbox{\small c}}$ & $N_{col}^{\mbox{\small d}}$ & $T^{\mbox{\small e}}$ & Density    & Composition \\
   Designation & $10^{16}$ cm   & M$_{\odot}$ & $10^{10}$cm$^{-3}$ & $10^{24}$cm$^{-2}$& 10$^3$ K & profile  &       \\
  \hline
   ``wall'' & 1-1.005 & 5  & 9.5 & 4.75 & 12 & uniform & WBH $^{\mbox{\small f}}$ \\
   ``thin layer''  & 1-1.2   & 10 & 3.32-4.76 &  7.93 & 3 & windlike & WBH \\
   ``thick layer'' & 0.5-1.1 & 10 & 0.237 & 14 & 3 & uniform & WBH \\
    ``ball''    & 0.1-1.1 & 2  & 0.016-1.84 & 16.7 & 10 & windlike & WBH \\
   ``cloud'' & 0.1-2   & 0.1 & 0.000357   & 0.0687 & 10 & uniform & solar \\
  \hline
    \multicolumn{8}{l}{}\\ [-3mm]
    \multicolumn{8}{p{15cm}}{$^{\mbox{\small a}}$ Outer and inner radii of the shell} \\
    \multicolumn{8}{p{15cm}}{$^{\mbox{\small b}}$ Total mass of the shell} \\
    \multicolumn{8}{p{15cm}}{$^{\mbox{\small c}}$ Baryonic number density} \\
    \multicolumn{8}{p{15cm}}{$^{\mbox{\small d}}$ Baryonic column density} \\
    \multicolumn{8}{p{15cm}}{$^{\mbox{\small e}}$ Initial temperature of uniform shells or interior layers of windlike shells} \\
    \multicolumn{8}{p{15cm}}{$^{\mbox{\small f}}$ Composition is taken from Woosley et al. (2007) (WBH stands for Woosley, Blinnikov, Heger)} \\
  \end{tabular}
  \end{table}

\clearpage

\hoffset=-20mm
\centerline {\bf ILLUSTRATIONS}


\begin{figure}[h]
\includegraphics[width=9cm]{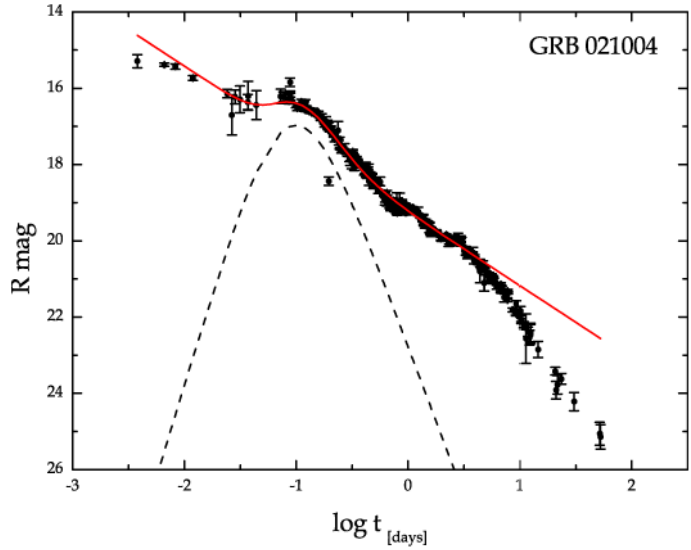}

\caption{The typical appearance of the afterglow lightcurve ``irregularity'' (GRB021004, Holland et al. (2003)).}
\end{figure}

 \vspace{2cm}

\begin{figure}[h]
 \includegraphics[width=12cm]{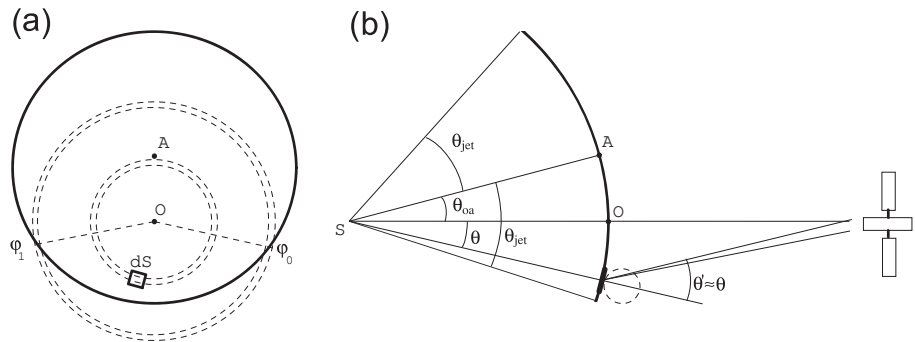}

\caption{Schematic pictures of the emitting region as it is seen ``from observer's point'' (a) and  in section(b), bold line depicts shell's outer surface. The GRB progenitor is in point S. SA is a jet axis, SO -- line of sight.}
\end{figure}

 \clearpage


\begin{figure}[t]
 \includegraphics[width=9cm]{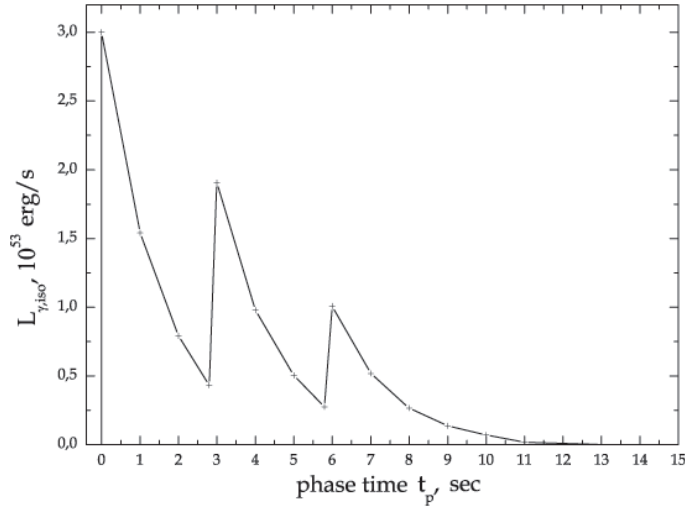}

\caption{Lightcurve of the incoming prompt emission.}
\end{figure}


\begin{figure}[b]
 \includegraphics[width=16cm]{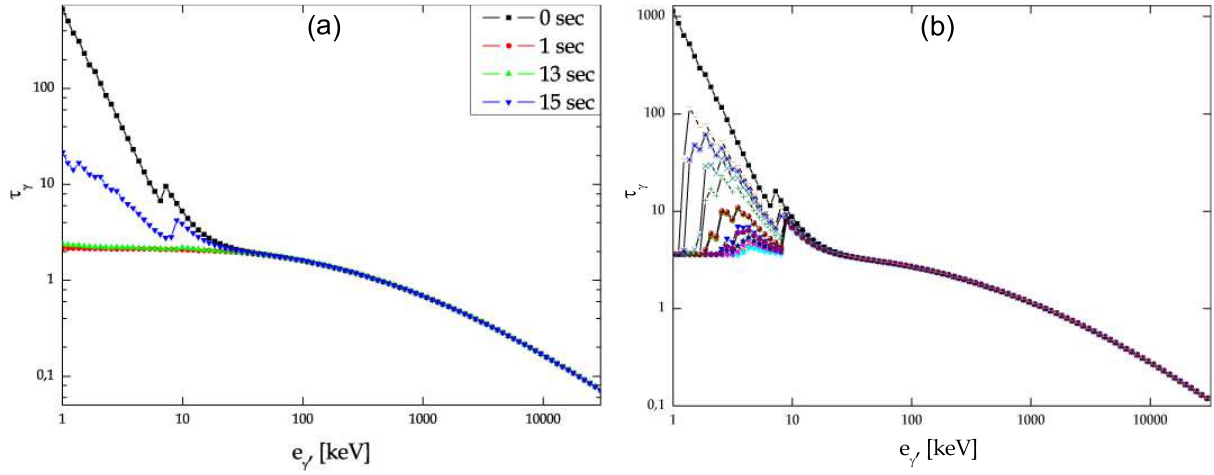}

\caption{Optical depths of shells at different phase time moments. Panel (a) -- the ``wall'' model with the system (12) used to calculate the state of matter; panel (b) -- one of early calculations for the ``thin layer'' with Boltzmann-Saha approximation used, where photoionization and Compton scattering affect only the heating power. It is clearly seen that in the latter case the temperature is insufficient to fully ionize heavy elements, but when the photoionization is ``turned on'', almost all electrons  instantly become ``blown away'' from their atoms by the radiation. And only at the last stages, when the gamma-ray photon density fades considerably, a partial recombination takes place.}
\end{figure}

 \clearpage

\begin{figure}[h]
 \includegraphics[width=16cm]{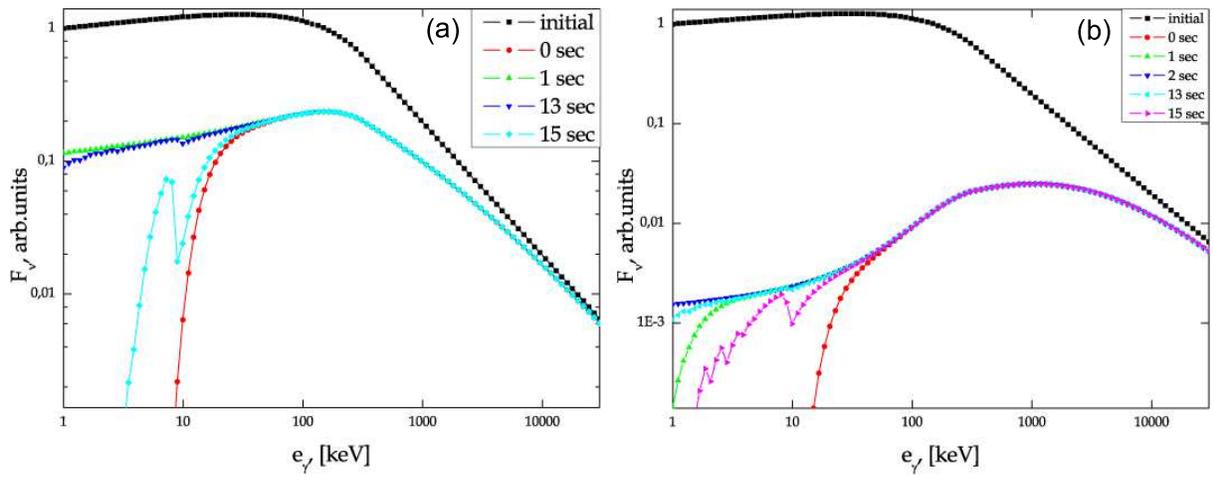}

\caption{Spectral shape of the gamma-ray emission coming into (initial) and passed through the shell at different phase times for models ``wall''(a) and ``thick layer''(b).}
\end{figure}

 \vspace{2cm}


\begin{figure}[h]
 \includegraphics[width=10cm]{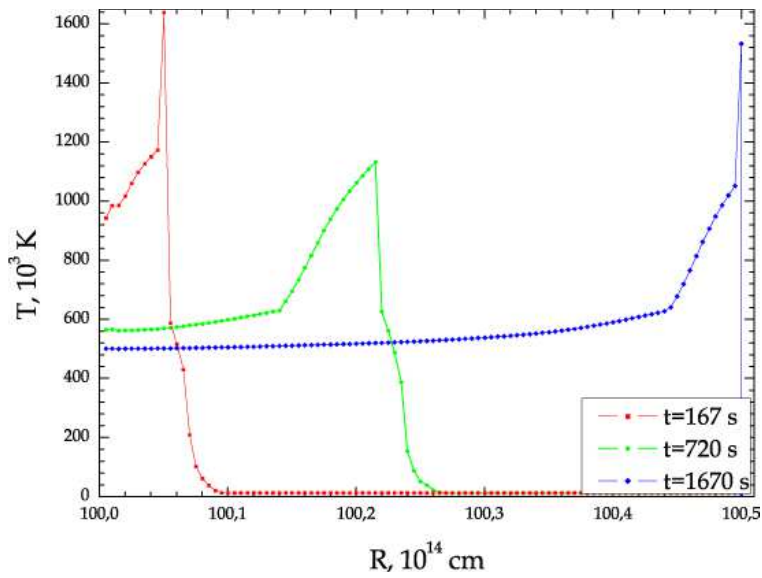}

\caption{Temperature profiles in the ``wall'' model (provides the best spatial resolution, 16.67 light seconds) taken in different times. }
\end{figure}

 \clearpage

\begin{figure}[h]
 \includegraphics[width=16cm]{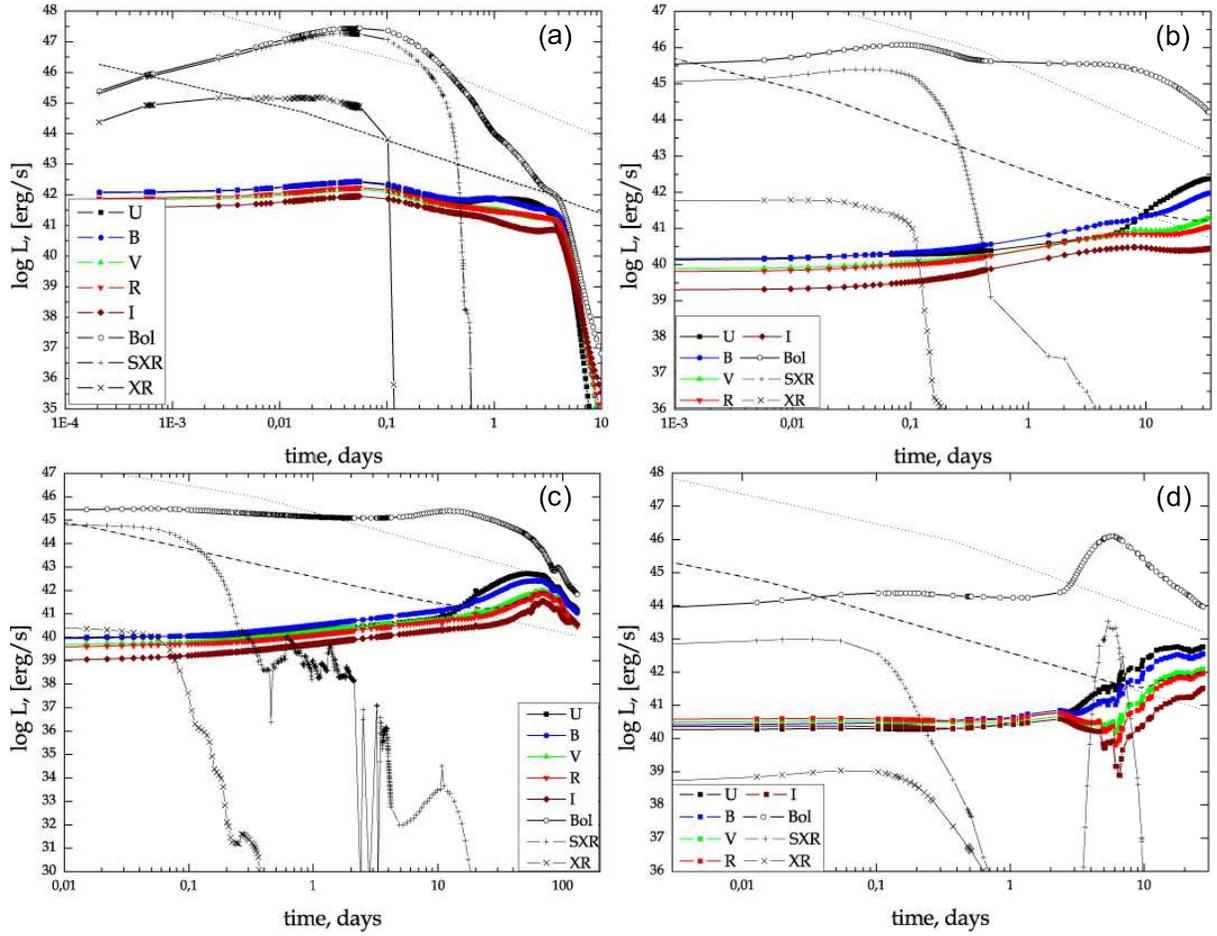}

\caption{Optical (UBVRI), X-ray (SXR for 0.1-2 keV, XR for 2-10 keV) and ``bolometric'' (1-50000 {\AA}) luminosity lightcurves for the following shell models: ``wall'' (a), ``thin layer''(b), ``thick layer''(c), ``ball''(d). The time is counted with respect to the prompt emission start. The subtle dotted lines represent the synchrotron power law afterglow characteristic region in the proper system R-filter. The dashed lines depict the sum of the lowest edge of the region and the thermal emission in R-band.}
\end{figure}

 \clearpage

\begin{figure}[h]
 \includegraphics[width=16cm]{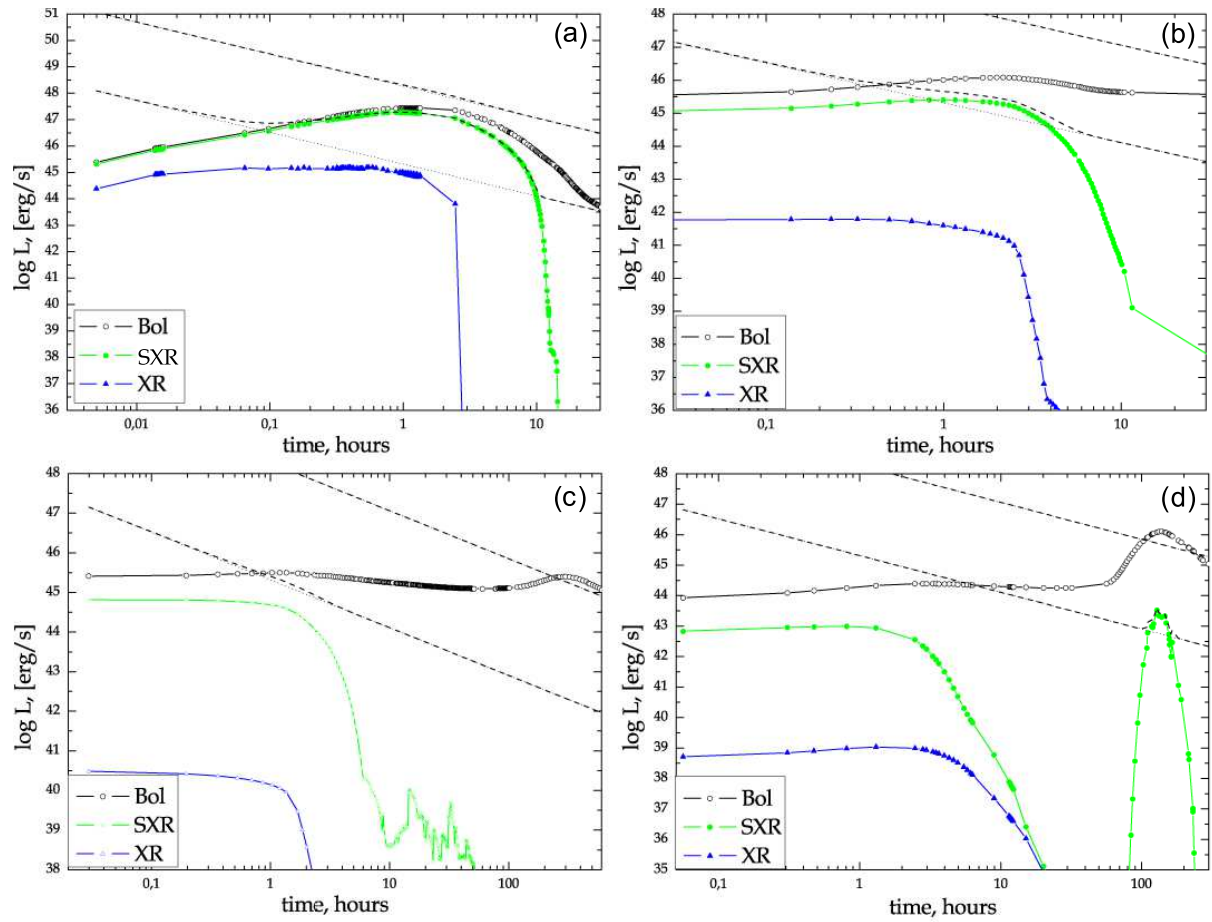}

\caption{X-ray (2-10 keV, XR), soft X-ray (0.1-2 keV, SXR) and ``bolometric'' luminosity lightcurves for the following shell models: ``wall'' (a), ``thin layer''(b), ``thick layer''(c), ``ball''(d). The subtle dotted lines represent the synchrotron power law afterglow characteristic region in the proper system 1-30 keV energy range. The dashed lines depict the sums of the region edges and the 0.1-10 keV thermal emission.}
\end{figure}

 \clearpage

\begin{figure}[h]
 \includegraphics[width=9cm]{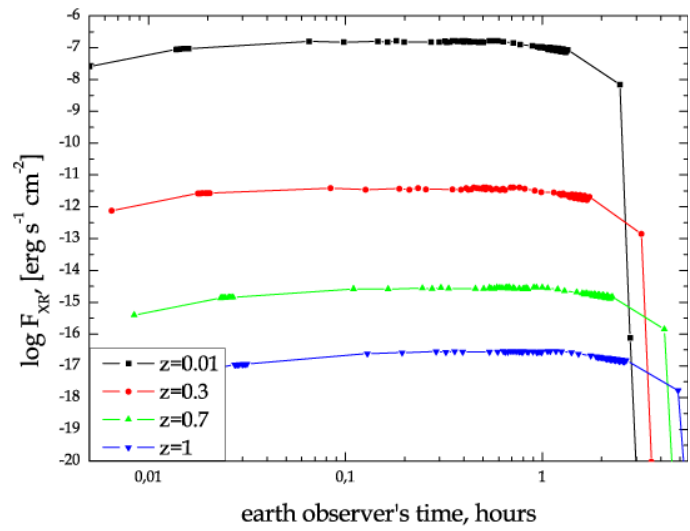}

\caption{X-ray 2-10 keV flux lightcurves as they seen from different redshifts. No extinction taken into account.}
\end{figure}

 \clearpage

\begin{figure}[h]
 \includegraphics[width=16cm]{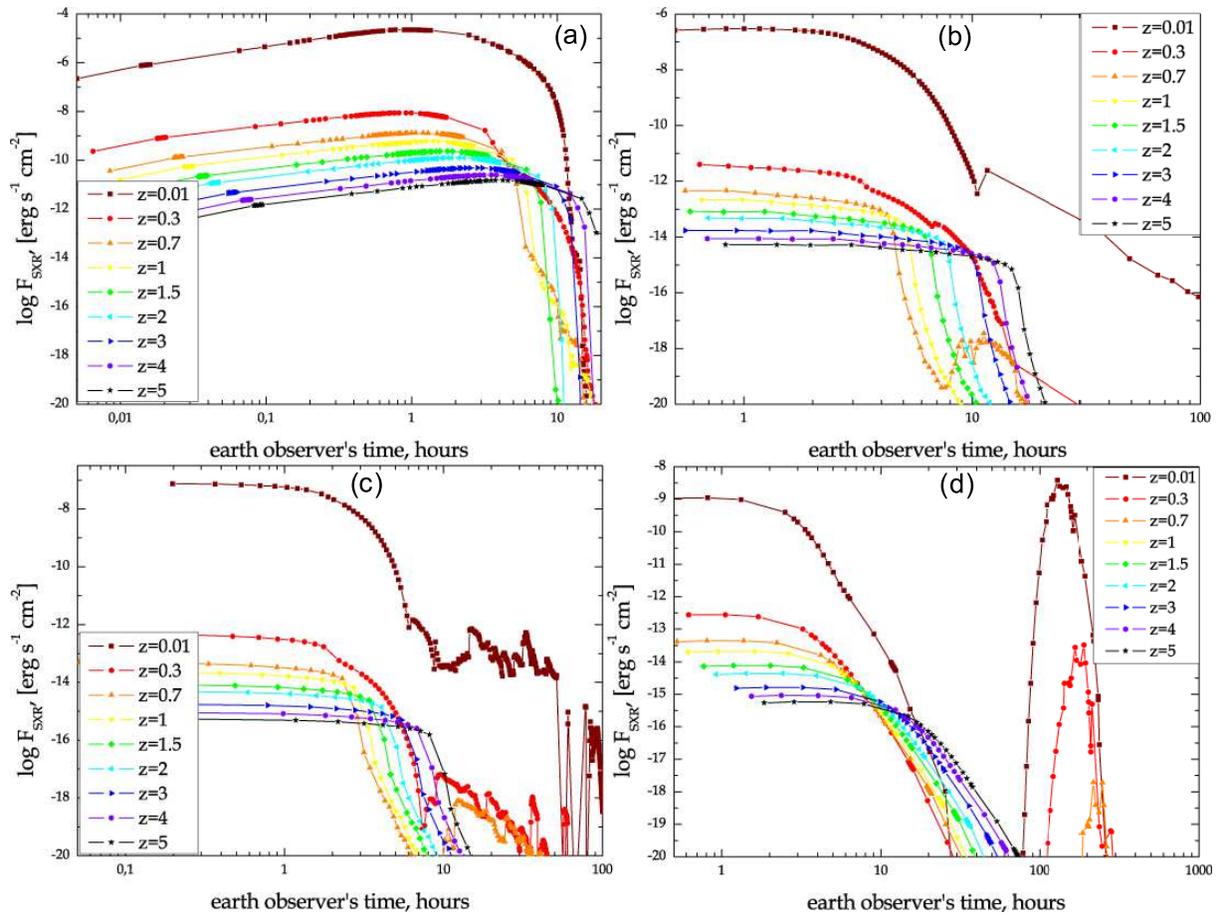}

\caption{Soft X-ray 0.1-2 kev flux lightcurves for the following shell models: ``wall'' (a), ``thin layer''(b), ``thick layer''(c), ``ball''(d), as they seen from different redshifts. No extinction taken into account.}
\end{figure}

 \clearpage

\begin{figure}[h]
 \includegraphics[width=16cm]{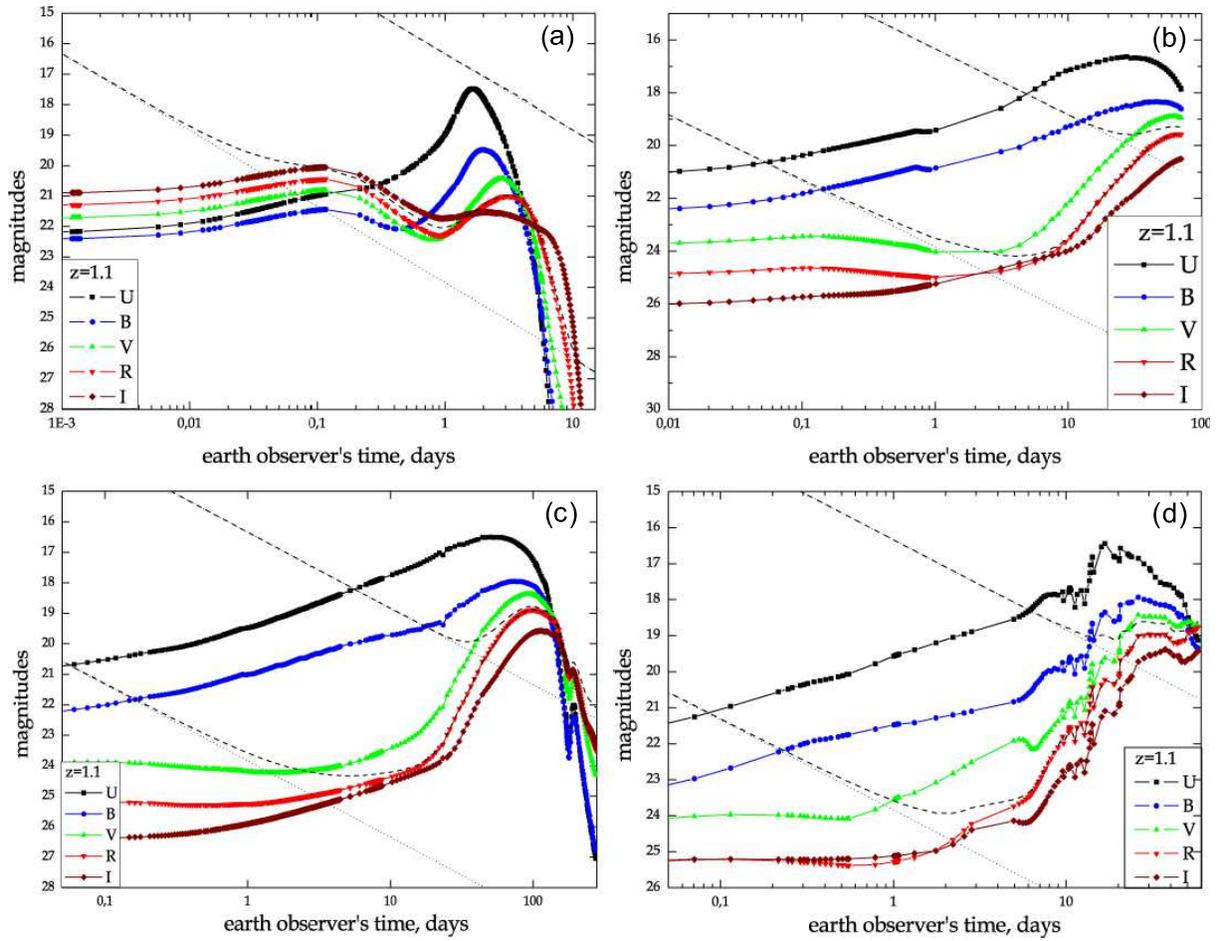}

\caption{Optical (UBVRI) lightcurves for the following shell models: ``wall'' (a), ``thin layer''(b), ``thick layer''(c), ``ball''(d), as they seen from redshift z=1.1. The subtle dotted lines represent the synchrotron power law afterglow characteristic region in observer's R-filter (i.e. it is somewhat different from those in GRB's proper system). The dashed lines depict the sum of the region edges and the thermal emission in R-band.}
\end{figure}

 \clearpage

\begin{figure}[h]
 \includegraphics[width=16cm]{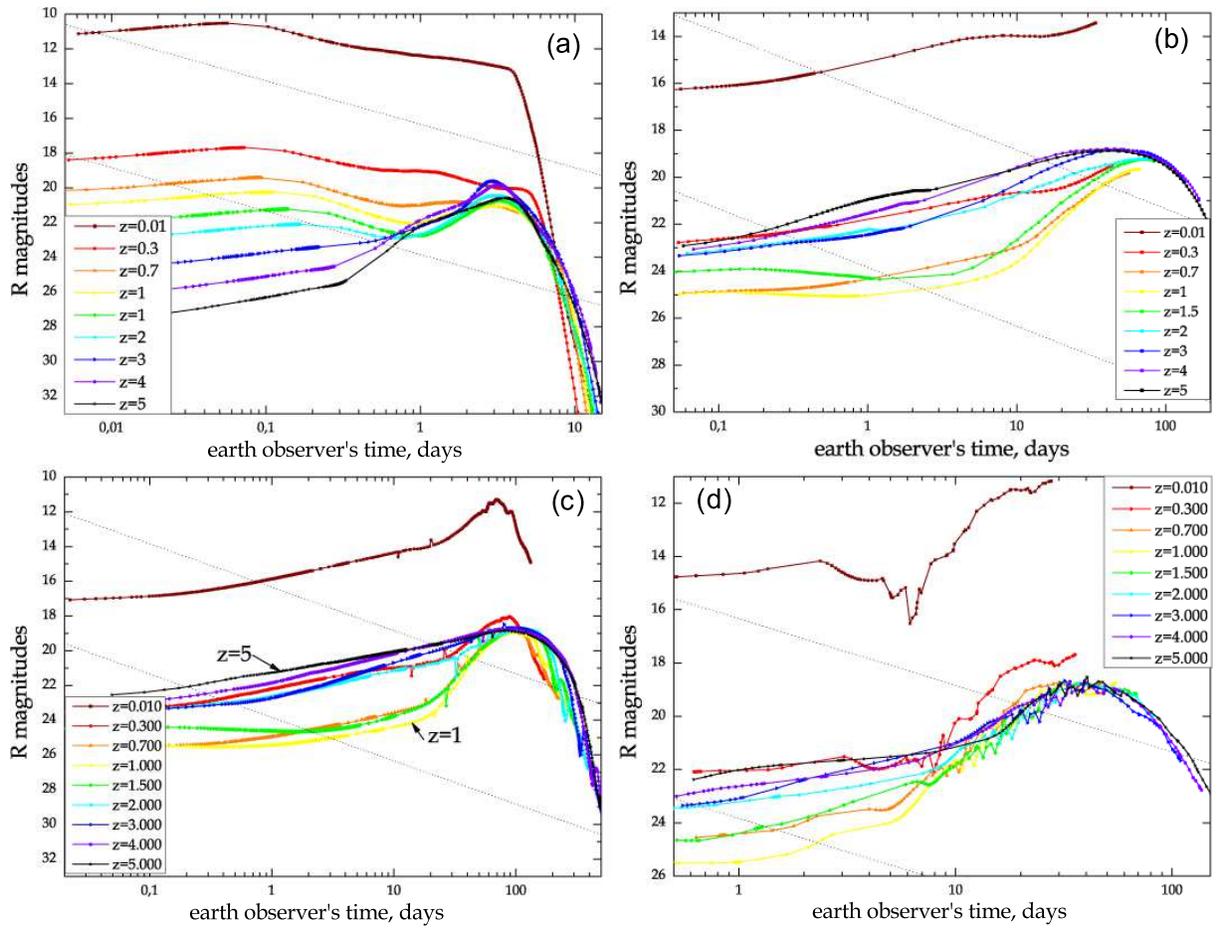}

\caption{R-magnitude lightcurves for the following shell models: ``wall'' (a), ``thin layer''(b), ``thick layer''(c), ``ball''(d), as they seen from different redshifts. The subtle dotted lines represent the synchrotron power law afterglow characteristic region in observer's R-filter. }
\end{figure}

 \clearpage

\begin{figure}[h]
 \includegraphics[width=16cm]{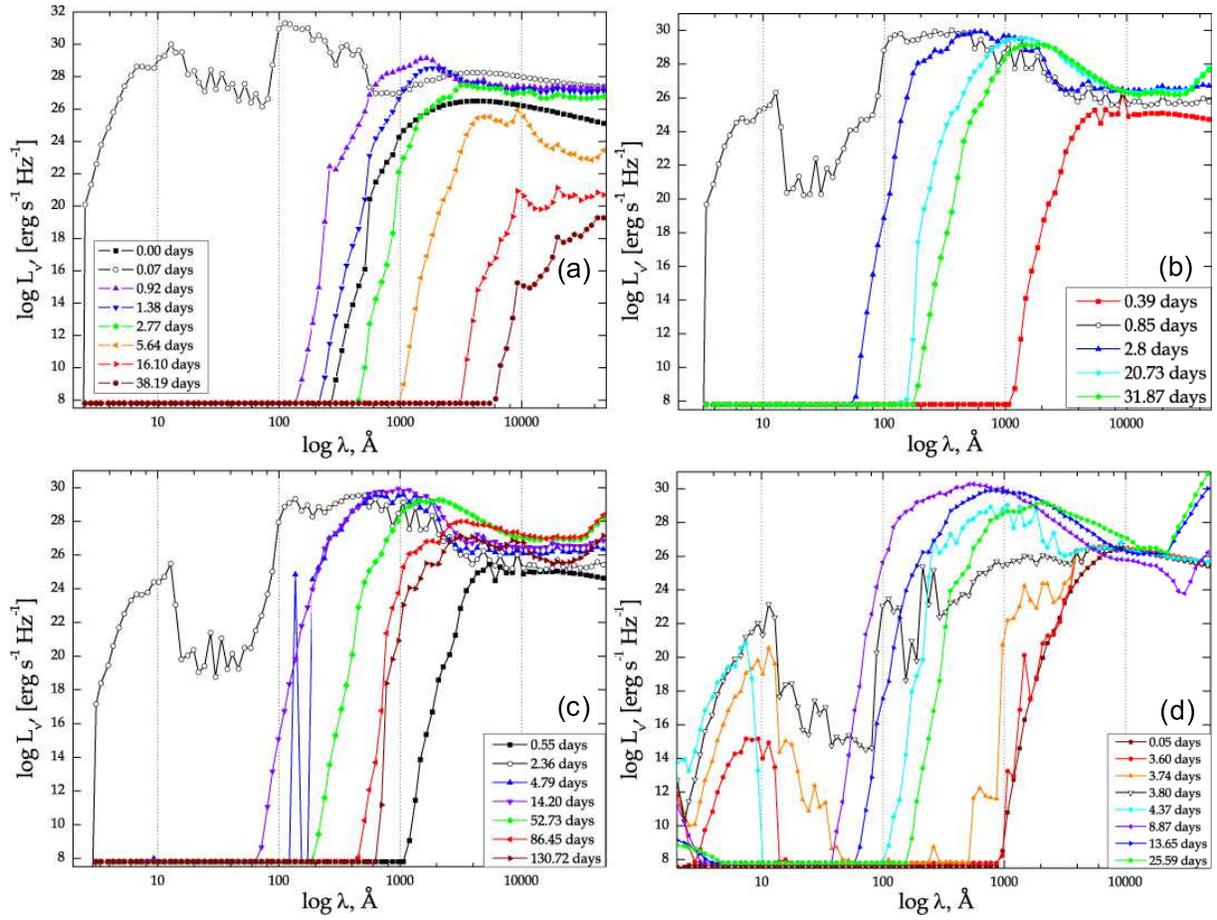}

\caption{Luminosity spectral density of the thermal emission at different proper system time moments for the following shell models: ``wall'' (a), ``thin layer''(b), ``thick layer''(c), ``ball''(d).}
\end{figure}

 \clearpage

\begin{figure}[h]
 \includegraphics[width=16cm]{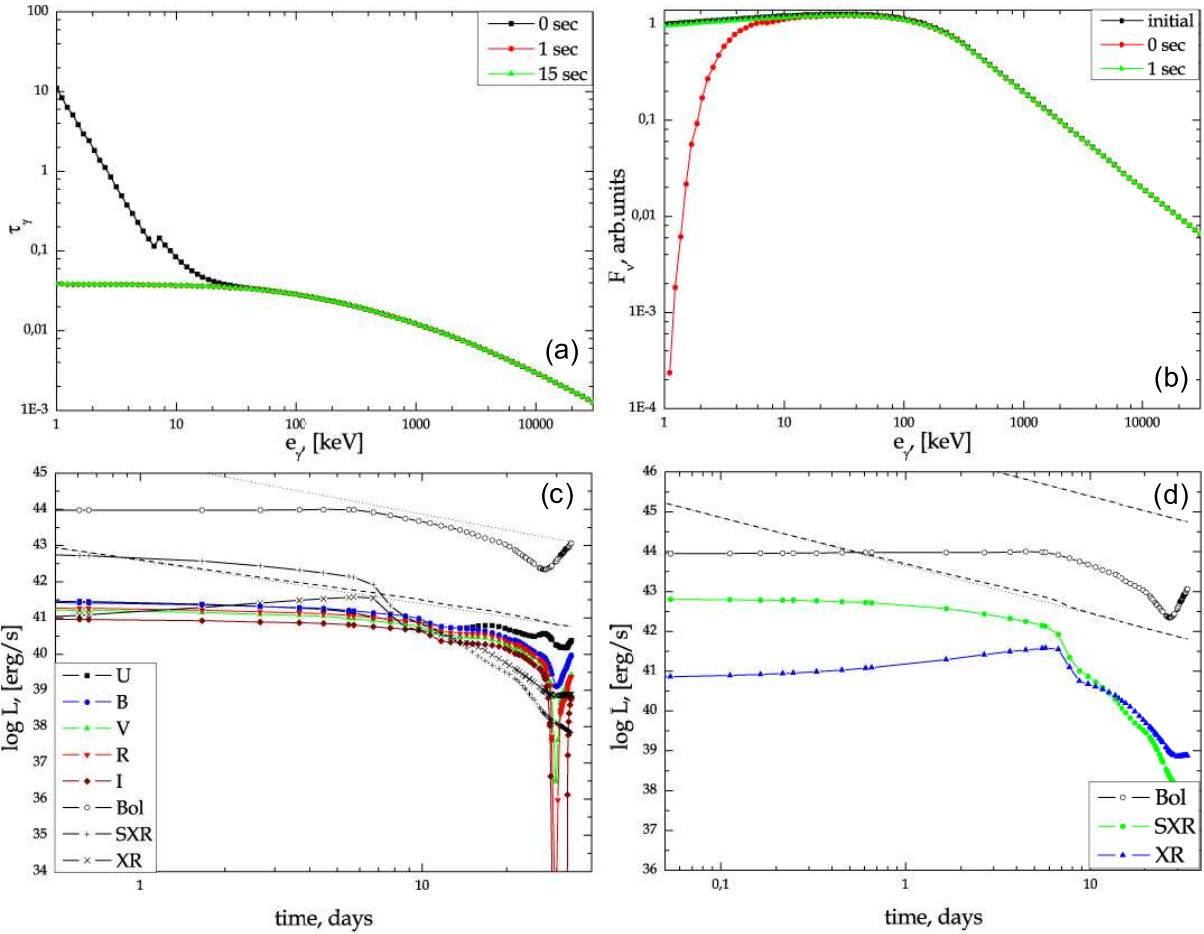}

\caption{Optical depths (a), gamma-ray spectra (b), luminosity lightcurves (c), X-ray luminosity lightcurves (d) for the ``cloud'' model. }
\end{figure}

 \clearpage

\begin{figure}[h]
 \includegraphics[width=16cm]{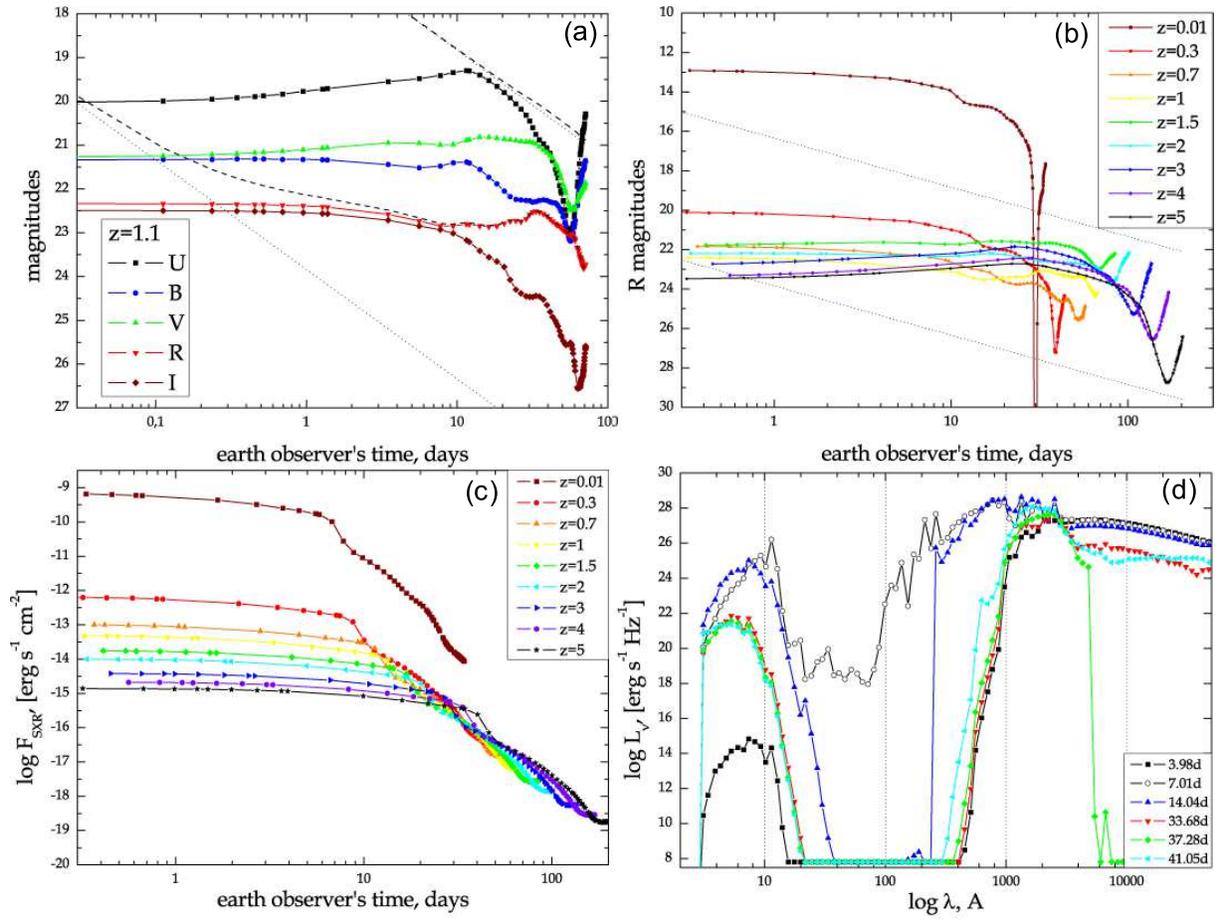}

\caption{Optical magnitude (a, b) and soft x-ray flux lightcurves, as they seen from different redshifts, and thermal emission spectra (d) for the ``cloud'' model.
}
\end{figure}

 \clearpage

\begin{figure}[h]
 \includegraphics[width=16cm]{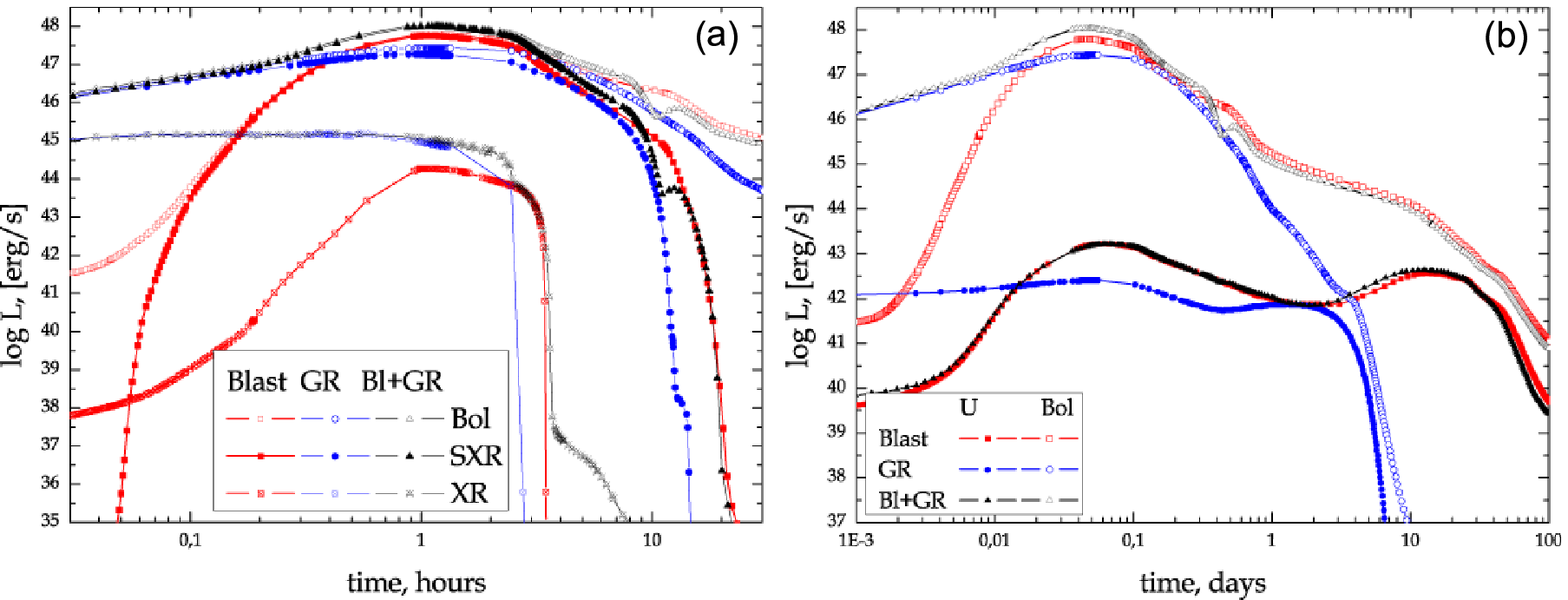}

\caption{X-ray (a) and optical (b) lightcurves for the ``wall'' model and different kinds of heating: radiative only (GR), kinetic only (Bl for blast) and combined (GR+Bl).}
\end{figure}

 \clearpage

\begin{references}
  \reference{\it Badjin et al.} (D.A. Badjin, G.M. Beskin, G. Greco), Astronomy Letters {\bf35}, 7 (2009).
  \reference{\it Band et al.}(D. Band, J. Matteson, L. Ford \etal), \apj\ {\bf413}, 281 (1993).
  \reference{\it Barkov \& Bisnovatyi-Kogan} ( M.V. Barkov, G.S. Bisnovatyi-Kogan), Astronomy Reports, {\bf49}, 611 (2005).
  \reference{\it Bisnovatyi-Kogan \& Timokhin} (G.S. Bisnovatyi-Kogan, A.N. Timokhin), Astronomy Reports, {\bf41}, 423 (1997).
  \reference{\it Bj\"{o}rnsson et al.} (G. Bj\"{o}rnsson, E.H. Gudmundsson, G.J\'{o}hannesson), \apj\ {\bf615}, L77 (2004).
  \reference{\it Blinnikov et al.} (S.I. Blinnikov, R. Eastman, O.S. Bartunov \etal), \apj\ {\bf496}, 454 (1998).
  \reference{\it Costa et al.} (E. Costa, F. Frontera, J. Heise), \nat\ {\bf387}, 783 (1997).
  \reference{\it Gehrels et al.} (N. Gehrels, E. Ramirez-Ruiz, D. B. Fox) arXiv:0909.1531v1 (2009).
  \reference{\it Holland et al.} (S.T. Holland, M. Weidinger, J.P.U. Fynbo \etal), \apj\ {\bf125}, 2991 (2003).
  \reference{\it Lazzati et al.} (D. Lazzati, E. Rossi, S. Covino \etal), \aap\ {\bf396}, L5 (2002).
  \reference{\it M\'{e}sz\'{a}ros} (P.M\'{e}sz\'{a}ros), \araa\ {\bf40}, 137 (2002).
  \reference{\it van Paradijs et al.} (J. van Paradijs, P.J. Groot, T. Galama \etal), \nat\, {\bf386}, 686 (1997).
  \reference{\it Piran} (T. Piran), Rev.Mod.Phys. {\bf76} 1143 (2004).
  \reference{\it Postnov} (K.A. Postnov), Phys.-Usp. {\bf42}, 469 (1999).
  \reference{\it Postnov et al.} (K.A. Postnov, S.I. Blinnikov, D.I. Kosenko, E.I. Sorokina), Nucl. Physics B (Proc. Suppl.), {\bf132} 327 (2004).
  \reference{\it Sazonov et al.} (S.Yu.Sazonov, J.P. Ostriker, R.A. Sunyaev), \mnras\ {\bf347}, 144 (2003).
  \reference{\it Shen et al.} (R.-F. Shen, P. Kumar, T. Piran),{\it To be published in the Mon. Not. Roy. Astron. Soc.}, preprint astro-ph.HE arXiv:0910.5727 (2009).
  \reference{\it Verner \& Yakovlev} (D.A. Verner, D.G. Yakovlev), \aaps\ {\bf109}, 125 (1995).
  \reference{\it Verner et al.} (D.A. Verner, F.J. Ferland, K.T. Korista, D.G. Yakovlev), \apj\ {\bf465}, 487 (1996).
  \reference{\it Woosley \& Bloom} (S.E. Woosley, J.S. Bloom), \araa\ {\bf44}, 507 (2006).
  \reference{\it Woosley et al.} (S.E. Woosley, S.I. Blinnikov, A. Heger) \nat\ {\bf450}, 390 (2007).
  \reference{\it Zhang \& MacFadyen} (W. Zhang, A. MacFadyen), \apj\ {\bf698}, 1261 (2009).

\end{references}
\end{document}